\input harvmac
\input epsf.tex
\overfullrule=0mm
\newcount\figno
\figno=0
\def\fig#1#2#3{
\par\begingroup\parindent=0pt\leftskip=1cm\rightskip=1cm\parindent=0pt
\baselineskip=11pt
\global\advance\figno by 1
\midinsert
\epsfxsize=#3
\centerline{\epsfbox{#2}}
\vskip 12pt
{\bf Fig.\the\figno:} #1\par
\endinsert\endgroup\par
}
\def\figlabel#1{\xdef#1{\the\figno}}
\def\encadremath#1{\vbox{\hrule\hbox{\vrule\kern8pt\vbox{\kern8pt
\hbox{$\displaystyle #1$}\kern8pt}
\kern8pt\vrule}\hrule}}
\def\appendix#1#2{\global\meqno=1\global\subsecno=0\xdef\secsym{\hbox{#1.}}
\bigbreak\bigskip\noindent{\bf Appendix. #2}\message{(#1. #2)}
\writetoca{Appendix {#2}}\par\nobreak\medskip\nobreak}

\Title{T97/036, SU-4240-660}
{{\vbox {
\medskip
\centerline{{\bf Effects of Self-Avoidance on the Tubular Phase of}}
\centerline{{\bf Anisotropic Membranes}}
}}}

\bigskip
\centerline{M. Bowick\footnote{}{\kern -20pt email: bowick@npac.syr.edu, 
guitter@spht.saclay.cea.fr},}
\medskip
\centerline{\it Physics Department, Syracuse University,}
\centerline{\it Syracuse NY 13244-1130, USA}
\medskip
\centerline{E. Guitter}
\medskip
\centerline{ \it CEA, Service de Physique Th\'eorique de Saclay,}
\centerline{ \it F-91191 Gif sur Yvette Cedex, France}
\baselineskip=12pt
\vskip .5in
 
We study the tubular phase of self-avoiding anisotropic membranes.
We discuss the renormalizability of the model Hamiltonian describing this
phase and derive from a renormalization group equation some general 
scaling relations for the exponents of the model. 
We show how particular choices of renormalization factors reproduce 
the Gaussian result, the Flory theory and the Gaussian Variational treatment 
of the problem. We then study the perturbative renormalization to one loop
in the self-avoiding parameter using dimensional regularization  
and an $\epsilon$-expansion about the upper critical dimension, 
and determine the critical exponents to first order in $\epsilon$.
\noindent
\vfill\eject
\nref\JERU{``Statistical Mechanics of Membranes and Surfaces,'' D.R. Nelson, 
T. Piran and S. Weinberg eds, Proceedings of the fifth Jerusalem Winter 
School for Theoretical Physics (World Scientific, Singapore, 1989).}
\nref\LESH{``Fluctuating Geometries in Statistical Mechanics and Field
Theory,'' F. David, P. Ginsparg and J. Zinn-Justin eds; Les Houches
Session LXII (Elsevier Science, The Netherlands, 1996) 
(http://xxx.lanl.gov/lh94).}
\nref\RT{L. Radzihovsky and J. Toner, Phys. Rev. Lett. {\bf 75} (1995) 4752
(cond-mat/9510172); University of Colorado preprint (cond-mat/9708046).}
\nref\KKN{Y. Kantor, M. Kardar and D.R. Nelson, Phys. Rev. Lett. {\bf 57} 
(1986) 791, Phys. Rev.  {\bf A 35} (1987) 3056.}
\nref\NP{D.R. Nelson and L. Peliti, J. Phys. France {\bf 48} (1987) 1085.}
\nref\KN{Y. Kantor and D.R. Nelson, Phys. Rev. Lett. {\bf 58} (1987) 2774;
Phys. Rev. {\bf A 36} (1987) 4020.}
\nref\PKN{M. Paczuski, M. Kardar and D.R. Nelson, Phys. Rev. Lett. {\bf 60} 
(1988) 2638.}
\nref\DG{F. David and E. Guitter, Europhys. Lett. {\bf 5} (1988) 709.}
\nref\GDLP{E. Guitter, F. David, S. Leibler and L. Peliti, Phys. Rev. Lett. 
{\bf 61} (1988) 2949; J. Phys. France {\bf 50} (1989) 1787.}
\nref\BFT{M. Bowick, M. Falcioni and G. Thorleifsson, Phys. Rev. Lett. 
{\bf 79} (1997) 885 (cond-mat/9705059).}
\nref\GRE{G. Grest, J. Phys. I France {\bf 1} (1991) 1695.}
\nref\SAM{M. Kardar and D.R. Nelson, Phys. Rev. Lett. {\bf 58} (1987) 
1289, 2280(E), Phys. Rev. {\bf A 38} (1988) 966.}
\nref\AL{J.A. Aronovitz and T.C. Lubensky, Europhys. Lett. {\bf 4} (1987) 395.}
\nref\DU{B. Duplantier, Phys. Rev. Lett. {\bf 58} (1987) 2733.}
\nref\DDG{F. David, B. Duplantier and E. Guitter, Phys. Rev. Lett. {\bf 72}
(1994) 311 (cond-mat/9307059); cond-mat/9702136.}
\nref\DW{F. David and K. Wiese, Phys. Rev. Lett. {\bf 76} (1996) 4564
(cond-mat/9602125); Nucl. Phys. {\bf B 487} (1997) 529 (cond-mat/9608022).}


\newsec{Introduction}

The statistical mechanics of isotropic tethered membranes has been 
extensively studied [\xref\JERU,\xref\LESH]. 
In a recent paper Radzihovsky and Toner (RT) \RT\
showed that intrinsically anisotropic tethered membranes are surprisingly
rich systems. In particular they exhibit an intermediate 
tubule phase between the crumpled and flat phases typical 
of isotropic tethered membranes [\xref\KKN-\xref\GDLP]. 
The tubule phase is characterized 
by being extended in one direction and crumpled in the other.
Furthermore any degree of anisotropy is expected to be relevant,
so such systems could be widespread in nature and very important.   
It is not hard to imagine many situations in which the polymerization of
a fluid membrane occurs anisotropically.

Recently the existence of this tubule phase for physical
anisotropic membranes has been confirmed by large-scale Monte Carlo 
simulations \BFT\ and the crumpled-to-tubule and tubule-to-flat
phase transitions both observed. In the case of self-avoiding
tethered membranes, current numerical evidence suggests that
the crumpled phase is destroyed in physical dimensions \GRE.
This enhances the possible significance of an ordered tubule phase
for self-avoiding anisotropic physical membranes {---} the only
transition left in this case may be the tubule-to-flat transition.
 
In this paper we study the effects of self-avoidance in the tubule
model of a self-avoiding tubule, previously introduced and
analyzed by RT \RT. This model may be considered as the analog 
of the Edwards model of self-avoiding membranes [\xref\SAM-\xref\DU],  
appropriately adapted to the tubular geometry, with bending rigidity in 
the extended direction of the tubule and self-avoidance in its crumpled
direction.

In Section 2 we use a renormalization group equation to reproduce some
of the critical exponent scaling relations of RT and derive some new
ones. These relations hold provided that the bending energy term
is not renormalized, and imply that there is only
one independent exponent in the model. Special cases of this treatment
reproduce the trivial Gaussian model as well as the Flory theory and
the Gaussian Variational approximation results of RT.

In Section 3 we establish the perturbative renormalizability
of the model and prove that the bending energy term is indeed
not renormalized. 

In Section 4 we calculate the critical exponents to first order
in an $\epsilon$-expansion about the upper critical dimension
for the relevance of self-avoidance. We use the techniques of dimensional
regularization and the Multi-local Operator Product Expansion (MOPE)
of reference \DDG .
We give the corresponding predictions of all relevant critical exponents
for the case of a physical membrane in the tubule phase.

\newsec{Scaling Relations}

\fig{(a) An anisotropic membrane with a stiff $y$ direction and a soft
$x_\perp$ direction; (b) after embedding, the membrane forms a tubule,
extended in the stiff $y$ direction and crumpled in the soft direction.
}{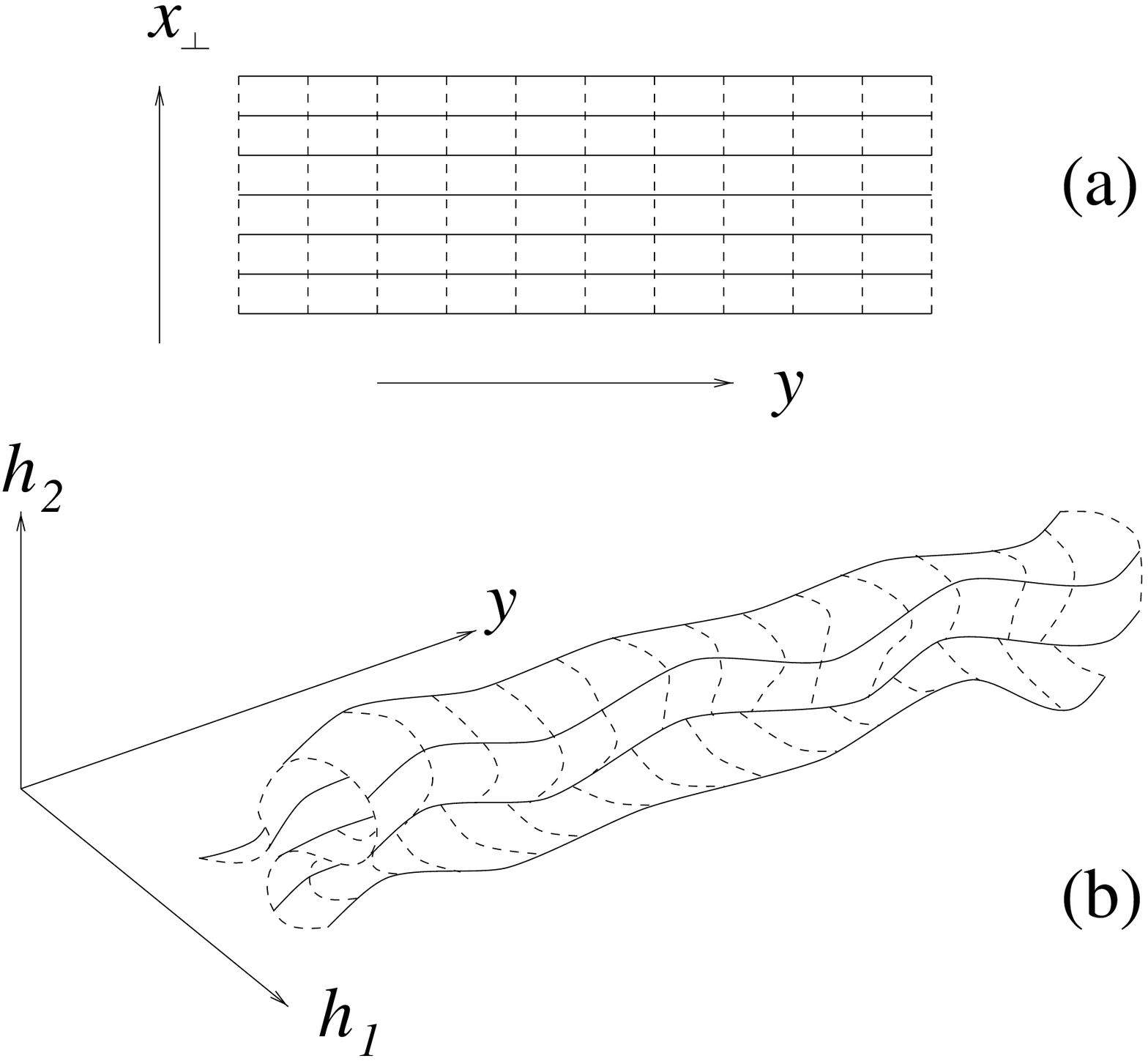}{8,truecm}
\figlabel\tube
We start by reviewing RT's model and scaling results 
for the tubular phase of self-avoiding anisotropic membranes \RT. 
We consider the generalized 
case of $D$-dimensional objects ($D=2$ corresponding to membranes) with
one stiff direction $y$ and $(D-1)$ soft directions $x_\perp$ (see
Fig. \tube-(a)). In the tubular phase, such an object will be extended 
in the $y$-direction and crumpled in the transverse direction. 
Using a Monge-like representation, the point with coordinates 
$(x_\perp,y)$ in the membrane will occupy a position 
$(\vec h(x_\perp,y),y)$ in the $d$-dimensional embedding space, 
with ${\vec h}$ a $(d-1)$-dimensional vector field perpendicular 
to the $y$ direction.
Adapting the Edwards model for self-avoiding membranes to the geometry
of the tubular phase RT obtained the Hamiltonian \RT\ 
\eqn\hamilt{\eqalign{\CH={1\over 2} & \int d^{D-1}x_\perp \, dy \ \left\{
\left( \partial^2_y {\vec h}(x_\perp,y)\right)^2+\left(\partial_\perp 
{\vec h}(x_\perp,y) \right)^2\right\}\cr & + {b\over 2} \int  d^{D-1}
x_\perp \, d^{D-1}x^\prime_\perp \, dy \ \delta^{(d-1)}\left({\vec h} 
(x_\perp,y)-{\vec h}(x^\prime_\perp,y) \right)\ .\cr}}
The first two terms describe the elastic properties of the 
membrane in the absence of self-avoidance, and represent a bending
energy term in the extended stiff $y$ direction and an effective 
entropically generated elastic term in the crumpled direction.
The third term is a two-body contact interaction with excluded
volume (or self-avoiding) parameter $b$. Due to the extended nature
of the tubule in the $y$ direction, the self-avoiding interaction
involves only points which have the same $y$ coordinate along the 
membrane \RT.
\medskip
The engineering dimensions of the fields and coordinates are
$[y]=1$, $[x_\perp]=2$ and 
\eqn\barezeta{\zeta_0\equiv[\vec h]= {5\over 2}-D\ .} 
This implies $[b]=-\epsilon$ with \RT\ 
\eqn\eps{\epsilon = 3D - {1\over 2} - \left({5\over 2}-D\right)d.}
We consider the model for $3/2<D<5/2$ only, where the bare roughness
exponent $\zeta_0$ \barezeta\ satisfies $0<\zeta_0<1$. Setting
$\epsilon=0$ fixes the upper critical dimension for the
relevance of the self-avoiding interaction to be \RT\
\eqn\duc{d_{uc}(D)={6D-1\over 5-2D}\ ,}
with, in particular, $d_{uc}(2)=11$.

\medskip
In Section 3 we will show that the Hamiltonian \hamilt\ renormalizes onto
itself. In other words, one can find renormalization factors $Z$, $Z_\perp$
and $Z_b$ such that the renormalized theory, using the renormalized Hamiltonian 
\eqn\Rhamilt{\eqalign{\CH^R={1\over 2} & \int d^{D-1}x^R_\perp \, dy \ 
\left\{Z \left( \partial^2_y {\vec h^R}(x^R_\perp,y)\right)^2+Z_\perp 
\left(\partial^R_\perp {\vec h^R} (x^R_\perp,y) \right)^2\right\}\cr & 
+ {Z_b b^R \mu^\epsilon \over 2} \int  d^{D-1}x^R_\perp \, 
d^{D-1}x^{\prime R}_\perp \, dy \ \delta^{(d-1)}\left({\vec h^R} 
(x^R_\perp,y)-{\vec h^R}(x^{\prime R}_\perp,y) \right) , \cr}} 
gives finite results at $\epsilon=0$ when expressed in terms of 
the renormalized self-avoiding parameter $b^R$.
We will moreover show that the bending energy term is not renormalized,
that is 
\eqn\Zone{Z=1.}
This assumption is crucial in the derivation of the scaling laws below.
The bare \hamilt\ and renormalized \Rhamilt\ Hamiltonians can be made
identical by appropriate rescalings of the height field ${\vec h^R}$, 
transverse coordinate $x^R_\perp$ and self-avoiding coupling $b^R$
in the following way: 
\eqn\Rquant{\eqalign{{\vec h^R}(x^R_\perp,y)&=Z_\perp^{(1-D)/4} 
{\vec h}(x_\perp,y)\cr x^R_\perp&=Z_\perp^{1/2} x_\perp \cr
b^R&=b\mu^{-\epsilon} Z_b^{-1}Z_\perp^{(1-D)(d+3)/4} . \cr}}
Consider now the height fluctuations in the bare model as determined
by the correlation function 
\eqn\propnu{G(x_\perp,y)\equiv -{1\over 2(d-1)}\left\langle 
\left({\vec h}(x_\perp,y)- {\vec h}(0,0)\right)^2\right\rangle . }
{}From \Rquant\ the renormalized version of this correlation function  
satisfies
\eqn\propR{G^R(x^R_\perp,y)\equiv -{1\over 2(d-1)}\left\langle 
\left({\vec h^R}
(x^R_\perp,y)- {\vec h^R}(0,0)\right)^2\right\rangle_R=Z_\perp^{(1-D)/2}
G(x_\perp,y) .} 
Writing $\mu\left.{d\phantom{\mu}\over d\mu}\right\vert_0
\left(Z_\perp^{(D-1)/2}G_R(x^R_\perp,y)\right)=0$, where the derivative is 
taken at fixed bare parameter $b$, we get the renormalization group 
equation 
\eqn\rgeq{ \mu {\partial\phantom{\mu}\over \partial \mu} G^R 
+{1\over 2}\, \delta\, x_\perp  {\partial\phantom{x_\perp}\over \partial 
x_\perp} G^R +{D-1\over 2}\, \delta\, G^R=0 , }
where  $\delta=\mu\left.{d\phantom{\mu}\over d\mu}\right\vert_0 \log Z_\perp$.
We suppose here that an infrared stable fixed point is reached, describing
the large scale properties of the membrane. The equation \rgeq\ holds
precisely at this fixed point. 
On the other hand simple scaling gives the homogeneity equation
\eqn\scaling{ \mu {\partial\phantom{\mu}\over \partial \mu} G^R 
-y {\partial\phantom{y}\over \partial y} G^R
-2 x_\perp  {\partial\phantom{x_\perp}\over \partial x_\perp} G^R
+(5-2D) G^R=0 .}
We thus get the fixed point renormalization group equation
\eqn\anscaltwo{y {\partial\phantom{y}\over \partial y} G^R
+ {1 \over z} x_\perp  {\partial\phantom{x_\perp}\over 
\partial x_\perp} G^R - 2 \zeta G^R=0 , }
where the anisotropy exponent $z$ and the roughness exponent $\zeta$
are given by
\eqn\expo{\eqalign{z&={2 \over 4 + \delta}\cr \zeta&= \zeta_0 + 
{ 1-D \over 4} \delta\ ,}}
with the bare roughness exponent $\zeta_0$ given by \barezeta .
Eq.~\anscaltwo\ implies the scaling
\eqn\anscalthree{G_R(x^R_\perp,y) \sim y^{2\zeta} 
F_1(y/(x^R_{\perp})^{z}) \sim (x^R_{\perp})^{2\nu} F_2(y/
(x^R_{\perp})^{z}),}
where the size exponent $\nu$ and roughness exponent $\zeta$
are related by $\nu=\zeta z$. Eliminating $\delta$ in \expo , 
we thus find the very general scaling relations 
\eqn\znexp{\eqalign{\zeta &= {3\over 2} + {1-D \over 2z}\cr
\nu&={3z\over 2} +{(1-D)\over 2}\ .\cr}}
Rewriting Eq.~\anscalthree\ in momentum space and using the derived
scaling relations one finds that the inverse of the height field
propagator $\tilde G^{-1}(q,p_{\perp})$ scales as 
$\tilde G^{-1}(q,p_{\perp}) = q^4 f(q/{p_{\perp}}^z)$.
Thus the anomalous dimension, $\eta$, for the bending rigidity
vanishes, as required by the non-renormalization theorem $Z=1$.
Similarly it is simple to show that  
$\tilde G^{-1}(q,p_{\perp}) = {p_{\perp}}^{2 + \eta_{\perp}} 
g(q/{p_{\perp}}^z)$, with $\eta_{\perp} = 4z - 2$. Since the
size exponent $\nu$ must exceed its phantom value
$\zeta_0/2$, one has $z > 1/2 $ and therefore $\eta_{\perp} > 0$.

{}From the above scaling relations, we end up with only one independent
exponent in the theory, depending on the precise value of $\delta$.
This value, and the subsequent predictions for all exponents, may be fixed
by imposing one more constraint on the renormalization factors
of our model Hamiltonian. At this stage, this extra imposed constraint
is totally arbitrary and different constraints lead to different
values of the exponents. It is interesting nevertheless to explore 
limiting cases where scaling is dominated by one component of the
Hamiltonian only, either the elastic term or the self-avoiding 
interaction. The corresponding limiting values of the exponents
indeed define the range of values in which the exact exponents are 
expected to fall. One can fix the scaling from the elastic terms 
only by assuming the absence of renormalization for the 
${\vec h}$ field, i.e. by imposing $Z_\perp=1$, yielding $\delta=0$. 
One then recovers the bare values $\zeta=\zeta_0$, $z=1/2$, 
$\nu=\zeta_0/2$ of the Gaussian theory without self-avoidance.
On the other hand we can consider the strong coupling limit where 
scaling is fixed by the self-avoiding term only. This yields the 
Gaussian Variational result, as discussed in Section 2.2 below and
also treated in \RT. 
A third, intermediate, estimate of the exponents is the Flory result, 
obtained by balancing the elastic and self-avoiding contributions 
in the Hamiltonian, as discussed in Section 2.1 and in \RT. 
Notice finally that 
these different estimates become exact and identical on the $\epsilon=0$ 
line, and can be used as starting points for a systematic expansion 
in the $(D,d)$ plane around this line. This idea has been used in
\DW\ for the self-avoiding isotropic membrane Edwards model.
In Section 4, we will calculate the correction to the
Gaussian, Flory and Variational estimates of the roughness exponent $\zeta$, 
at first order in $\epsilon$ and for fixed $D=2$. 

\subsec{Flory Theory}

In Flory theory one assumes that elastic energies are
comparable to self-avoiding energies. If this is to remain true
under renormalization one should require that both terms
renormalize in the same way viz.\ $Z_\perp = Z_b$.
Given this assumption one finds from \Rquant\ that 

\eqn\bflory{b^R=b\mu^{-\epsilon} Z_\perp^{(1-D)(d+3)/4 - 1} .}
In this case the fixed point condition directly determines
$\delta$ in terms of $\epsilon$ to be
\eqn\deltaf{\delta_F = {-4\epsilon \over {\{ 4 + (D-1)(d+3)\}}}}
with $\epsilon$ as in \eps\ .
Using \expo\ and \znexp , the size exponent $\nu$ is then found to be 
\eqn\nuf{\nu_F = {(D+1) \over (d+1)}\ ,}
which coincides with the Flory prediction found in \RT .
This is nothing but the usual Flory result for a $(D-1)$--dimensional
self-avoiding object in a $(d-1)$--dimensional embedding space, and
corresponds to treating the different transverse slices of the tubule
as independent \RT.
The other exponents are likewise determined in this 
approximation to be 
\eqn\zf{\eqalign{z_F &=  {{4 + (D-1)(d+3)} \over 3(d+1)} \cr
		 \zeta_F &= {3(D+1) \over {4 + (D-1)(d+3)}}\ . } }
The corresponding values for the physical tubule ($D=2$ and $d=3$)
are $\delta_F = -8/5$, $\nu_F = 3/4$, $z_F=5/6$ and $\zeta_F=9/10$.

\subsec{Gaussian Variational Approximation}

A different approximation one can make is to assume that
the self-avoiding term is not renormalized viz.\ $Z_b=1$.
This is exactly the approximation which is made in 
a Gaussian Variational treatment of the problem, where 
the exact density functional is approximated  by the best possible
Gaussian weight for the field $\vec h$, using a variational
principle \RT. In this case the field $\vec h$ is renormalized,
but the self-avoiding interaction term is not.
Repeating the above analysis in the case $Z_b=1$, one finds easily
from the fixed point condition 
\eqn\deltav{\delta_{\rm var} = {-4\epsilon \over (D-1)(d+3)}\ .}
The size exponent $\nu$ in this approximation, first obtained by RT, is
\eqn\nuv{\nu_{\rm var} = {7(D-1) \over (3d-5)}\ ,}
and the other exponents are likewise determined to be 
\eqn\zv{\eqalign{z_{\rm var} &= {(D-1)(d+3) \over (3d-5)} \cr
	      \zeta_{\rm var}&= {7 \over (d+3)}\ . } }
The corresponding values for the physical tubule
are $\delta_{\rm var} = -2/3$, $\nu_{\rm var}=7/4$, $z_{\rm var}= 3/2$,
and $\zeta_{\rm var}=7/6$. 
The unphysical nature of these values ($\nu$ and $\zeta$ cannot 
exceed $1$) indicates that, in this approximation,
the tubule phase is unstable. For $D=2$, in fact, 
one sees from \nuv\ that the tubule phase in unstable below $d=4$.
It is known, however, that the Gaussian Variational method is a strong
coupling method which usually over-estimates the size exponent.

\newsec{Renormalizability}

We now turn to the issue of the perturbative renormalizability 
of the theory for $\epsilon \ge 0$. We rely on the general formalism 
introduced in reference \DDG\ for the treatment of non-local 
interactions. In the diagrams of the 
perturbative expansion in $b$ we first identify the singular configurations 
of interacting
points which contain possible divergences. We then use a short distance 
multi-local operator product expansion (MOPE) to analyze these 
singularities and show that they are proportional to the insertion 
of multi-local operators. A simple power counting argument allows us 
to extract from all singular configurations those which give rise to 
actual divergences. This, together with some symmetry arguments, 
singles out all the operators which require renormalization. 
{}From this analysis, we deduce that the Hamiltonian \hamilt\ renormalizes 
onto itself, according to Eq.~\Rhamilt\ and moreover, $Z=1$,
i.e. there is no renormalization of the bending energy term.
Our analysis will be presented for $D=2$, but it could be easily extended to 
the range $3/2<D<5/2$ where the roughness exponent 
$\zeta_0=(5-2D)/2$ satisfies $0<\zeta_0<1$.

Let us concentrate on the partition function ${\cal Z}$ associated with 
the Hamiltonian \hamilt\ at $D=2$:
\eqn\partfunc{{\cal Z}_b=\int\CD[{\vec h}(x,y)]
\,\exp(-\CH[{\vec h}])\ .}
It can be expanded in powers of $b$ according to
\eqn\Zpertexp{
{\cal Z}_b={\cal Z}_0\ \sum_{N=0}^\infty\,{(-b/2)^N\over N!}\,
\int\prod_{i=1}^Ndx_i\,dx^\prime_i\, dy_i
\left\langle \prod_{i=1}^N\delta^{(d-1)}\left({\vec h}(x_i,y_i)-
{\vec h}(x^\prime_i,y_i)\right) \right\rangle_0
\ ,
}
where ${\cal Z}_0$ is the partition function of the non self-avoiding 
($b=0$) theory
and $\left\langle (\cdots)\right\rangle_0$ denotes the corresponding Gaussian 
average
\eqn\evevo{
\left\langle(\cdots)\right\rangle_0\ =\ {1\over{\cal Z}_0}\,
\int \CD[{\vec h}(x,y)]\,\exp\left(
-{1\over 2} \int dx\, dy \ \left\{
\left( \partial^2_y {\vec h}(x_,y)\right)^2+\left(\partial_x 
{\vec h}(x,y) \right)^2\right\}
\right)(\cdots)
\ .
}
Each $\delta$ function in \Zpertexp\ can be written as
\eqn\exprep{
\delta^{(d-1)}\left({\vec h}(x_i,y_i)-{\vec h}(x^\prime_i,y_i)\right)\ =\
\int{d^{d-1}{\vec k}_{i}\over (2\pi)^{d-1}}
\,{\rm e}^{{\rm i} {\vec k}_{i}\cdot \left({\vec h}(x_i,y_i)-{\vec h}(x^\prime_i,y_i)
\right)}
}
and one is lead to evaluate the Gaussian average 
\eqn\gav{\eqalign{\left\langle\prod_{i=1}^N
{\rm e}^{{\rm i}{\vec k}_i\cdot\left({\vec h}(x_i,y_i)
-{\vec h}(x^\prime_i,y_i)\right)}\right\rangle_0 = 
\exp\Big(-{1\over 2}
\sum_{i,j=1}^N{\vec k}_i\cdot{\vec k}_j\big\{
& G_0(x_i-x_j,y_i-y_j) \cr
- & G_0(x^\prime_i-x_j,y_i-y_j) \cr
- & G_0(x_i-x^\prime_j,y_i-y_j) \cr
+ & G_0(x^\prime_i-x^\prime_j,y_i-y_j) \big\} \Big)
\ , \cr}
}
where $G_0$ is the two-point function\foot{Here ${\rm erf}(u)$ denotes
the usual error function ${\rm erf}(u)\equiv (2/\sqrt{\pi})\int_0^udt 
\exp(-t^2)$.}
\eqn\propbis{\eqalign{G_0(x,y)&\equiv -{1\over 2(d-1)}\left\langle 
\left({\vec h}
(x,y)- {\vec h}(0,0)\right)^2\right\rangle_0\cr & = 
-{1\over 2\sqrt{\pi}}\, \vert x\vert^{1/2}\, \exp \left(
-{y^2\over 4 \vert x\vert }\right) -{1\over 4}\, y\, 
{\rm erf}\left({y\over 2\vert x\vert^{1/2}}\right) \ .\cr}}

\fig{The diagram of order $N$ in \Zpertexp\ is made of $N$ dipoles. The 
two end points of a given dipole are located at the same position $y_i$ in
the $y$ direction but at different positions $x_i$ and $x^\prime_i$ in the $x$ 
direction.  }{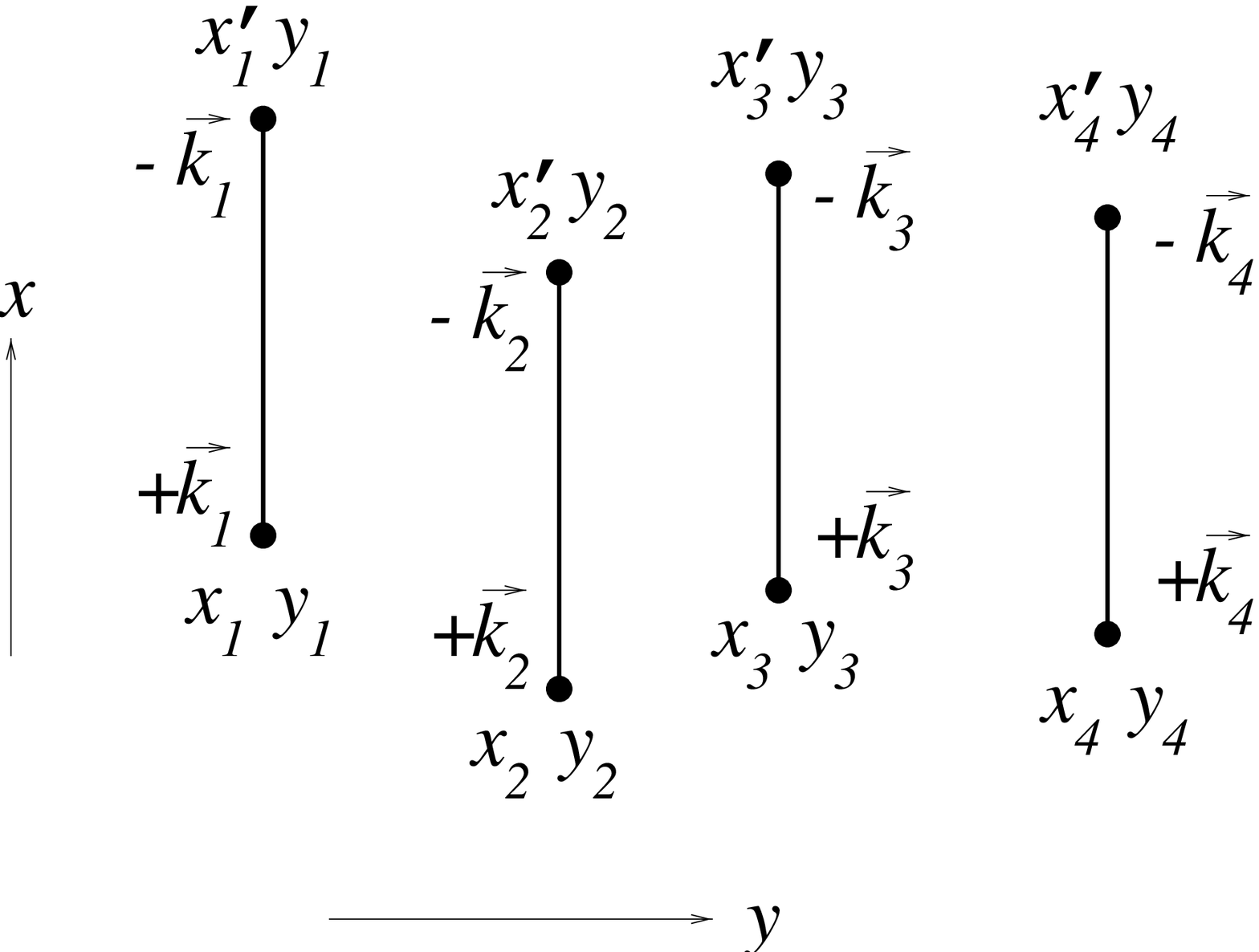}{8.truecm}
\figlabel\dipole
\noindent The term of order $N$ in the perturbative expansion \Zpertexp\ is therefore
naturally represented
by a diagram of $N$ ``dipoles" of interacting points located at 
$(x_i,y_i;x^\prime_i,y_i)$ with ``charge" $\pm {\vec k}_i$, as depicted 
in Fig.~\dipole . Note that the two end points of a given dipole $i$ are 
located at the same position $y_i$ in the $y$ direction but at different
positions $x_i$ and $x^\prime_i$ in the $x$ direction.
A singular configuration of interacting points is found
when the quadratic form 
\eqn\quadform{\eqalign{Q(\{{\vec k}_i\})=\sum_{i,j}{\vec k}_i\cdot
{\vec k}_j\big\{
G_0(x_i-x_j,y_i-y_j) 
- G_0(x^\prime_i-x_j,y_i-y_j)
& - G_0(x_i-x^\prime_j,y_i-y_j)  \cr
& + G_0(x^\prime_i-x^\prime_j,y_i-y_j)\big\} \cr}
}
appearing in \gav\ is not positive definite.
Using the integral representation of the two-point function
\eqn\twopoint{G_0(x,y)=\int {dp \over 2\pi} {dq\over 2\pi} {
{\rm e}^{{\rm i}\,(px+qy)}-1 \over q^4+p^2}\ ,}
we obtain
\eqn\quadint{Q(\{{\vec k}_i\})=\int {dp \over 2\pi} {dq\over 2\pi}{\left\vert
\sum_i{\vec k}_i\,{\rm e}^{{\rm i}qy_i}({\rm e}^{{\rm i}px_i}-
{\rm e}^{{\rm i}px^\prime_i})\right\vert^2 \over q^4+p^2}\ .}
The quadratic form $Q$ is thus positive definite except for those 
configurations of end points $\{x_i,x^\prime_i,y_i\}$ for which
one can find a set of charges $\{{\vec k}_i\}$, not all zero,
satisfying:
\eqn\zero{\eqalign{& 
\sum_i{\vec k}_i\,{\rm e}^{{\rm i}qy_i}({\rm e}^{{\rm i}px_i}-
{\rm e}^{{\rm i}px^\prime_i})={\vec 0}\quad \forall (p,q) \cr
\Leftrightarrow {\vec \rho}(x,y)&\equiv \sum_i{\vec k}_i\, 
\delta (y-y_i) \left( \delta(x-x_i)
-\delta(x-x^\prime_i)\right)={\vec 0} \quad \forall (x,y)\cr}}
This latter condition is the requirement that the charge density 
${\vec \rho}(x,y)$ vanishes identically, while some of the charges 
${\vec k}_i$ remain non-zero.
\fig{A molecule with 2 loops made of a connected assembly of 4 dipoles.
This molecule has 3 atoms located at different values of $x$ but at
the same value of $y$.}{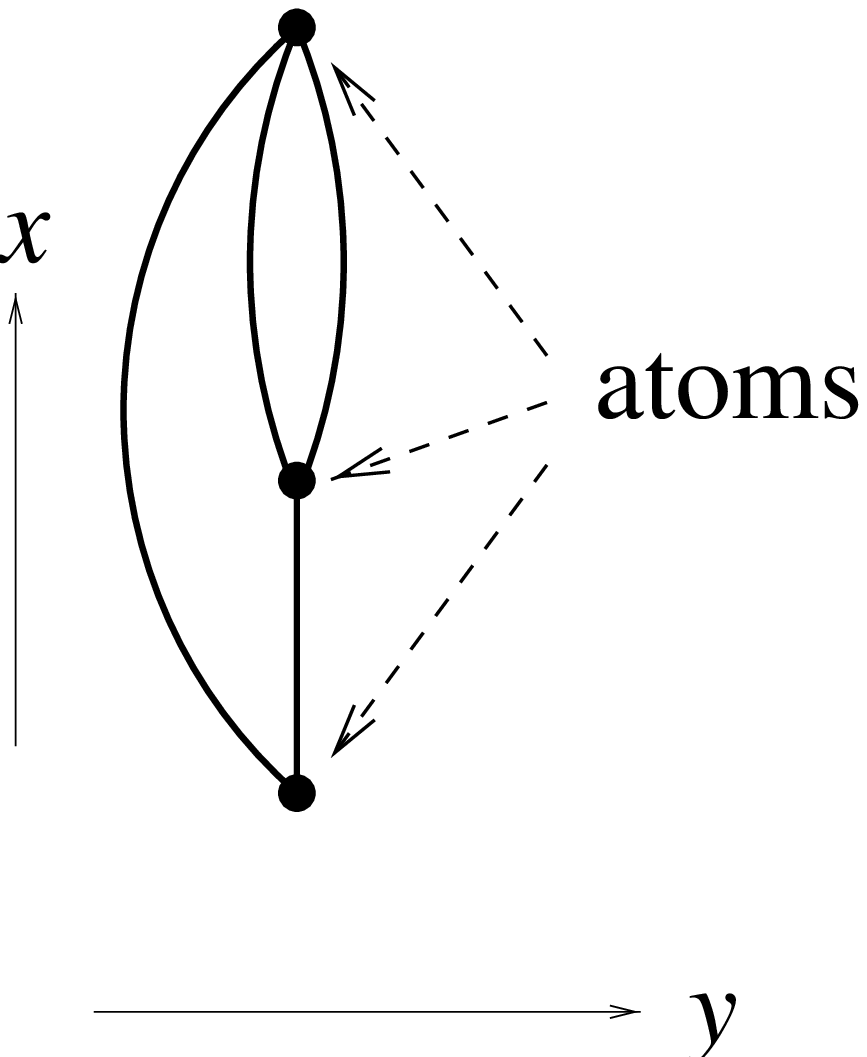}{5.truecm}
\figlabel\molecule
\noindent This is possible if some of the $N$ dipoles arrange to form 
a so-called ``molecule", i.e. attach their end points and assemble 
into a connected diagram with at least one loop, such as in 
Fig.~\molecule . A set of end points in contact form what is called
an ``atom" and their common position is the position of the atom. 
A zero of $Q$ is obtained by an appropriate choice of non-zero charges 
keeping all the atoms neutral, which is possible in the presence of a loop.
Note that all the atoms of a molecule have the same position in
the $y$ direction.
Note also that singularities coming from disconnected
molecules can be treated separately and that dipoles in the molecule
which do not belong to a loop (dead branches) do not contribute to
the singularity and can be ignored.  

The construction above identifies the singular configurations
of end points which give rise to possible divergences.
Such a configuration is characterized by a set of $M$ atoms 
labeled by $p$ and with position $(x_p,y)$ with the same $y$ coordinate
for all the atoms. For each
atom $p$, we denote by $I_p$ the set of dipoles $i$ which
attach their first end point at the atom $p$ (i.e. $(x_i,y_i)=(x_p,y)$),
and by $J_p$ the set of dipoles $j$ which attach their second
 end point at the atom $p$ (i.e. $(x^\prime_j,y_j)=(x_p,y)$).
The singularity can be analyzed by use of the general short distance
Multi-local Operator Product Expansion (MOPE) introduced in \DDG .
In practice, one can return to the operator level (left hand
side of \gav ) and write the contribution of the atom $p$ in \gav
\eqn\decomp{\eqalign{
\prod_{i\in I_p} {\rm e}^{{\rm i}{\vec k}_i\cdot {\vec h}(x_i,y_i)}
\prod_{j\in J_p} {\rm e}^{-{\rm i}{\vec k}_j\cdot {\vec h}
(x^\prime_j,y_j)}=&\left\langle \prod_{i\in I_p} {\rm e}^{{\rm i}
{\vec k}_i\cdot {\vec h}(x_i,y_i)} \prod_{j\in J_p} 
{\rm e}^{-{\rm i}{\vec k}_j\cdot {\vec h}(x^\prime_j,y_j)}
\right\rangle_0 \cr &\ \times : 
\prod_{i\in I_p} {\rm e}^{{\rm i}{\vec k}_i\cdot {\vec h}(x_i,y_i)}
\prod_{j\in J_p} {\rm e}^{-{\rm i}{\vec k}_j\cdot {\vec h}
(x^\prime_j,y_j)}
:\ ,\cr}}
i.e. separate in the right hand side of \gav\ the propagators 
$G_0$ which involve only points
inside the atom $p$, and which reconstruct precisely the Gaussian average
above, from those involving at least one end point not in the atom $p$, 
corresponding to a normal product prescription.  
This separation allows us to isolate the singularity in the 
factorized Gaussian average, while the normal product has a regular 
expansion
in $x_i-x_p$ ($i\in I_p$), $x^\prime_j-x_p$ ($j\in J_p$) and  $y_k-y$
($k\in I_p\cup J_p$)
\eqn\regexp{\eqalign{
:{\rm e}^{{\rm i}{\vec k}_p \cdot
{\vec h}(x_p,y)}\big 
(1+ &{\rm i} \sum_{i\in I_p} \left(
(x_i-x_p)\,{\vec k}_i\cdot \partial_x{\vec h}(x_p,y) +
(y_i-y)\,{\vec k}_i\cdot \partial_y{\vec h}(x_p,y) \right)\cr 
- &{\rm i} \sum_{j\in J_p} \left((x^\prime_j-x_p)\,{\vec k}_j\cdot 
\partial_x{\vec h}(x_p,y)+(y_j-y)\,{\vec k}_j\cdot 
\partial_y{\vec h}(x_p,y) \right)
+\ldots\big): \cr}}
with ${\vec k}_p=(\sum_{i\in I_p} {\vec k}_i-\sum_{j\in J_p}{\vec k}_j)$
being the total charge of the atom $p$. The same treatment can be applied to
all the atoms of the molecule, creating for each atom
$p$ an exponential factor ${\rm e}^{{\rm i}{\vec k}_p \cdot{\vec h}(x_p,y)}$,
together with insertions of various $\partial_x$ and/or $\partial_y$
derivatives of the field ${\vec h}$ at the point $(x_p,y)$. 
As in \DDG , the MOPE is obtained by performing the integration over
the charges ${\vec k}_i$ for the dipoles $i$ forming the molecule.
This expands the corresponding product of bi-local operators
$\prod_i\, \delta^{(d-1)}\left({\vec h}(x_i,y_i)-{\vec h}(x^\prime_i,y_i)
\right)$ around the chosen singular configuration in terms of general
$M$-body operators of the form
\eqn\genMO{\Phi(x_1,\ldots,x_M)=\int d^{d-1}{\vec h}
\,\prod_{p=1}^M\,A_p(x_p,y) \ \nabla_{\vec h}^{{\vec m}_p} \,
\delta^{(d-1)}\left({\vec h}-{\vec h}(x_p,y)
\right)\ ,}
multiplied by singular coefficients (see \DDG\ for details). 
Here $A_p(x_p,y)$ denotes either the unity operator $1$ or
a local operator in the derivatives of the field ${\vec h}$ at
point $(x_p,y)$ and $\nabla_{\vec h}^{\vec m}$ is a short hand notation
for $\prod_{\alpha=1}^{d-1} \partial^{m_\alpha}_{h_\alpha}$. 
The above operators are 
multi-local in the $x$ direction but local in the $y$ direction.
This is because all the atoms in the molecule have the same $y$-position. 
We will see two explicit examples of the MOPE in the next section
where explicit one-loop calculations are presented.

At this stage, let us mention the following important result
concerning the case where one inserted operator involves $\partial_y$
derivatives only (such as $(\partial_y^2{\vec h})^2$). Indeed,
such a term comes from the expansion of some operator
${\rm e}^{{\rm i}{\vec k}_i\cdot {\vec h}(x_i,y_i)}$
taken at $x_i=x_p$ exactly (we suppose here that $i\in I_p$ rather
than $i\in J_p$). However, in contrast with 
the coordinate $x_i$ (or resp. $x^\prime_i$)
which appears only in the atom $p$, the coordinate $y_i$ appears in a
second atom $p'$ (such that $i\in J_{p'}$), which is in general 
distinct from $p$. The expansion in $y_i-y$ can be done 
simultaneously on the operator ${\rm e}^{{\rm i}{\vec k}_i\cdot 
{\vec h}(x_i,y_i)}$ above for $x_i=x_p$ and for the operator
 ${\rm e}^{-{\rm i}{\vec k}_i\cdot {\vec h} (x^\prime_i,y_i)}$ 
for $x^\prime_i=x_{p'}$, in which case the operator to be expanded 
in $y_i-y$ is ${\rm e}^{{\rm i} {\vec k}_i\cdot ({\vec h}(x_p,y_i)
-{\vec h}(x_{p'},y_i))}$.
We thus get the important result that those operators with only 
partial derivatives in the $y$ direction can be regrouped so that
they involve the difference of the ${\vec h}$ field at two 
(in general different) points of the molecule.
An example of such operator is the two-body operator:
\eqn\exMO{\left(\partial_y({\vec h}(x_1,y)-{\vec h}(x_2,y))\right)^2
\,\delta^{(d-1)}\left({\vec h}(x_1,y)-{\vec h}(x_2,y)\right)} 
If the two end points of the dipole $i$ happen to belong to
the same atom $p$, then the operator to be expanded is $1$,
which means that this dipole cannot give rise to insertions of 
local operators with only $\partial_y$ derivatives.

This latter remark has an important implication for the renormalization 
of local operators, coming from the particular case of singular configurations 
where the molecule has only one atom. In this case, 
each dipole in the molecule falls automatically in the class just described
of dipoles with their two end points in the same,
unique, atom. We thus get the important result that local
operators with only $\partial_y$ derivatives 
are not created by renormalization. Such terms, when absent from
the original Hamiltonian \hamilt , never appear,
and the only such operator present in \hamilt\ (namely \
$(\partial_y^2{\vec h})^2$) is {\it not} renormalized, that is
\eqn\zonebis{Z=1\ .}

Having identified the singular configurations and the corresponding
general multi-local operators to which their singularities are 
proportional, it remains to identify those singularities which
are not integrable and give rise to actual divergences.
If the molecule is made of $K$ dipoles, the operator which
is expanded via the MOPE is the product of $K$ $\delta^{(d-1)}$ factors, 
with canonical dimension $-K \zeta_0 (d-1)$ in units of $y$.
The dimension of the multi-local operator $\Phi$ in \genMO\ is
\eqn\dimgen{(d-1)\zeta_0+\sum_{p=1}^M 
({\rm dim}[A_p] -\left(|{\vec m}_p|+ (d-1)\right)\zeta_0)\ ,}
with the notation $\vert{\vec m}\vert=\sum_{\alpha=1}^{d-1}m_\alpha$.
The corresponding singular coefficient in the MOPE has
thus dimension
\eqn\dimcoeff{-(K+1)(d-1)\zeta_0-\sum_{p=1}^M 
({\rm dim}[A_p] -\left(|{\vec m}_p|+ (d-1)\right)\zeta_0)}
This coefficient has to be integrated over the $2K-M$ relative 
$x$ coordinates and the $K-1$ relative $y$ coordinates of the
$2K$ end points of the dipoles approaching the positions of the 
$M$ atoms.
This gives a superficial degree of convergence for the corresponding
integral
\eqn\degsup{\eqalign{\omega&=2(2K-M)+(K-1)+\left(M-(K+1)\right)(d-1)
\zeta_0 +\sum_{p=1}^M(|{\vec m}_p|\zeta_0-{\rm dim}[A_p])\cr
&= 3(M -2) + \epsilon\, (K-M+1)
+\sum_{p=1}^M(|{\vec m}_p|\zeta_0-{\rm dim}[A_p])
\cr}}
with $\epsilon=5-(d-1)\zeta_0$. Note that $K-M+1$ is nothing but
the number of loops in the molecule.
A divergence is found whenever $\omega\le 0$.
It is easy to check that all the local operators $A$ but the unit 
operator have a strictly negative dimension in units of $y$, 
as a consequence of the relation 
$\zeta_0<1$.  At $\epsilon=0$, $\omega\le 0$ requires either $M=2$,
${\vec m}_1={\vec m}_2={\vec 0}$ and $A_1=A_2=1$, which
is nothing but the original contact interaction in \hamilt , or $M=1$,
in which case $\Phi$ is either the unity operator $1$ or a local 
operator $A(x,y)$ which moreover must satisfy ${\rm dim}[A]\ge -3$. 
We already know from the previous discussion that $A$ must contain
at least one $\partial_x$ derivative since terms with only
$\partial_y$ derivatives are not created. Due to the $x\to -x$
symmetry, the coefficient of a term with only one $\partial_x$
vanishes and one thus needs at least two $\partial_x$ derivatives. 
The term with largest dimension satisfying this criterion is the
original elastic term in \hamilt\ $(\partial_x{\vec h})^2$ 
which already has dimension $-3$. It is thus, together with
the unity operator, the only renormalized local
operator in the theory. The renormalization of the unity operator is simply
a shift in the free energy of the system. In particular, it disappears
in the computation of average values of physical observables, and can
be simply ignored.

In conclusion, we have shown that the Hamiltonian \hamilt\ renormalizes
onto itself, with $Z=1$, as announced.

For $\epsilon>0$, the theory is super-renormalizable since $\omega$
in \degsup\ increases with the number of loops of the molecule.

\newsec{One-loop calculations}

Let us now present one-loop calculations, which give the corrections
at first order in $\epsilon$ for the critical
exponents $\zeta$, $\nu$ and $z$. We use here dimensional
regularization by considering the theory at $\epsilon >0$ and 
by calculating the renormalization factors $Z_\perp$ and $Z_b$ 
needed to make the theory finite for $\epsilon=0$ at one-loop order 
in $b^R$. We use a minimal subtraction scheme where we keep  
for the first order correction in $Z_\perp$ and $Z_b$ only 
the corresponding pole in $\epsilon$. 
\fig{The two singular configurations 
leading to (a) a one loop renormalization of $(\partial_x{\vec h})^2$ 
and (b) a one loop renormalization of $\delta^{(d-1)}\left({\vec h}(x,y)
-{\vec h}(x^\prime,y)\right)$.}{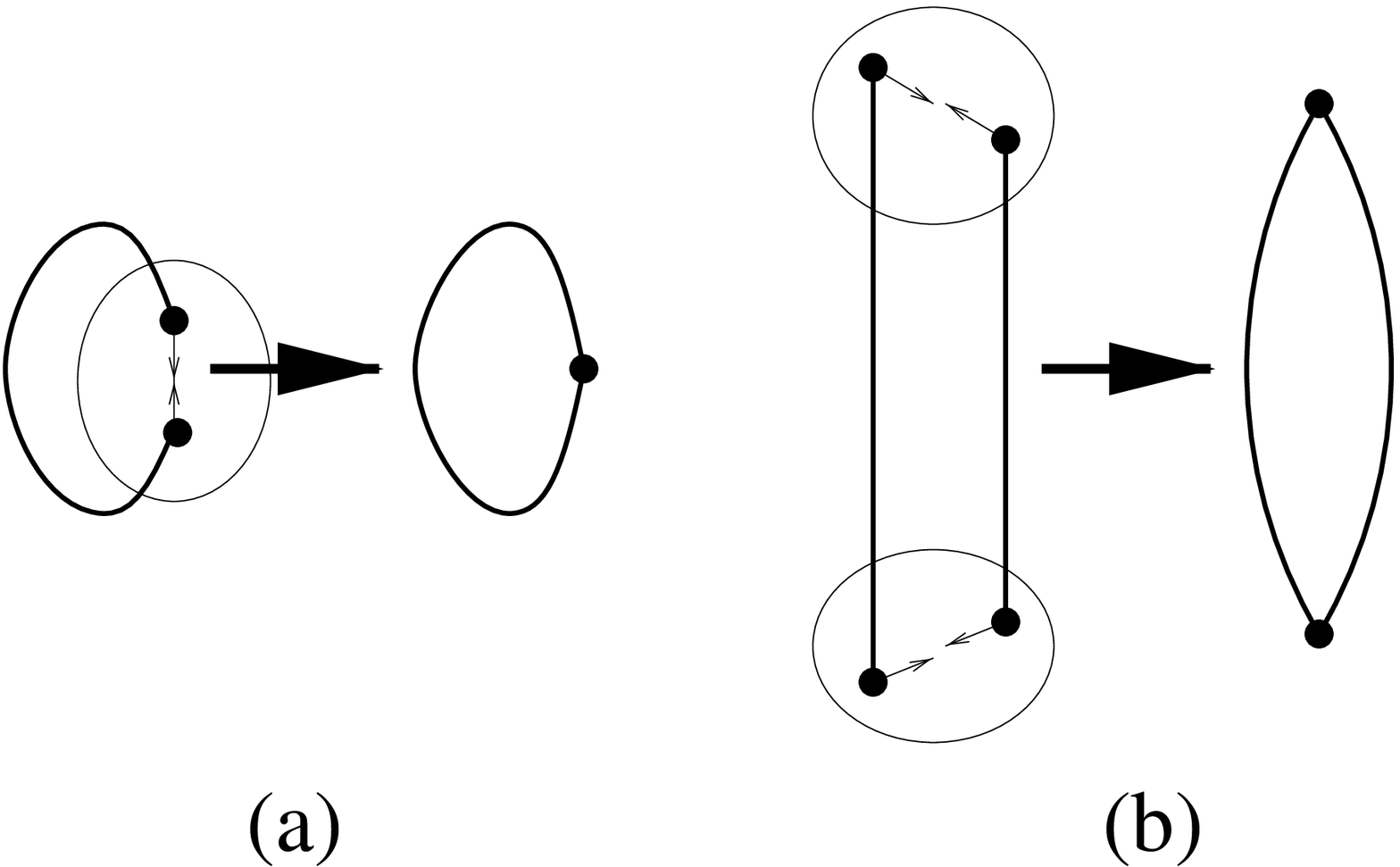}{8.truecm}
\figlabel\oneloop
We obtain our results in the framework of the
MOPE described above, which we use here in two simple cases:
the $1$ atom molecule made of a single dipole with its two
end points approaching each other (see Fig.~\oneloop-(a))
and the $2$-atom molecule made of two dipoles approaching
each other (see Fig.~\oneloop-(b)). 
The first situation will give us a renormalization of the local
operator $(\partial_x{\vec h})^2$, and hence a first order 
correction to $Z_\perp$.  The second situation will give us a 
renormalization of the bi-local operator 
$\delta^{(d-1)}\left({\vec h}(x,y)-{\vec h}(x^\prime,y)\right)$,
and hence a first order correction to $Z_b$.

To analyze the divergence for Fig.~\oneloop-(a),
we use the Operator Product Expansion (OPE): 
\eqn\OPEone{\eqalign{{\rm e}^{{\rm i}{\vec k} \cdot {\left({\vec h}(x,y)- 
{\vec h}(x^\prime,y)\right)}}={\rm e}^{{\vec k}^2\, G_0(x-x^\prime,0)}\Big\{
& 1+{\rm i}(x-x^\prime){\vec k}\cdot:\partial_x{\vec h}(x_0,y):\cr
&-{1\over 2}(x-x^\prime)^2 k_\alpha k_\beta:\partial_xh_\alpha(x_0,y)
\partial_xh_\beta(x_0,y):+\ldots\Big\}\cr}}
where $x_0=(x+x^\prime)/2$ and
\eqn\gzero{G_0(x-x^\prime,0)=-{1\over 2 \sqrt{\pi}} \vert x-x^\prime 
\vert^{1/2}\ .}
When integrated over ${\vec k}$, it gives the MOPE:
\eqn\MOPEone{\eqalign{\delta^{(d-1)}\left({\vec h}(x,y)-{\vec h}
(x^\prime,y)\right)= &{1\over (4\pi)^{{d-1\over 2}}}\Big\{
{1\over \left(-G_0(x-x^\prime,0)\right)^{{d-1\over 2}}}\times 1
\cr &
-{1\over 4}{(x-x^\prime)^2\over \left(-G_0(x-x^\prime,0)
\right)^{{d+1\over 2}}}\times : \left(\partial_x{\vec h}(x_0,y)
\right)^2 : +\ldots \Big\} \ .\cr}}
We use the above formula for the {\it renormalized} theory \Rhamilt\ , 
expanded to first order in $b^R$. As in \Zpertexp , the bi-local
$\delta$ interaction comes with a factor $-b^R\mu^\epsilon/2$ and
the singularity in \MOPEone\ proportional to 
$(\partial_x{\vec h^R})^2$ will
be cancelled by the corresponding counterterm, appearing with
a factor $-(Z_\perp-1)/2$, provided we choose:
\eqn\Zperp{{(Z_\perp-1)\over 2}= b^R {\mu^\epsilon\over 2} 
\int_{\vert X \vert \le \mu^{-2}}\kern -20pt dX \ 
{1\over 4}{(2\sqrt{\pi})^6\over (4\pi)^5}
{X^2\over (\vert X\vert^{1/2})^{{6-\epsilon}}}}
where $X=x^R-x^{\prime R}$, and where we have used 
\gzero\ and $d=11-2\,\epsilon$. 
This leads to:
\eqn\Zperpfinal{Z_\perp=1+{b^R\over 16\pi^2}{1\over \epsilon}\ .}

Let us now analyze the divergence for Fig.~\oneloop-(b).
We now use the OPE for the first atom
\eqn\OPEtwo{{\rm e}^{{\rm i}\left({\vec k}_1\cdot{\vec h}(x_1,y_1)
+{\vec k}_2\cdot{\vec h}(x_2,y_2)\right)}={\rm e}^{-{\vec k}_1\cdot
{\vec k}_2G_0(x_1-x_2,y_1-y_2)}:{\rm e}^{{\rm i}({\vec k}_1
+{\vec k_2})\cdot{\vec h}(x_0,y_0)}\{1+\ldots\}:}
where $x_0=(x_1+x_2)/2$ and $y_0=(y_1+y_2)/2$,
and the similar OPE for the second atom
\eqn\OPEtwobis{{\rm e}^{-{\rm i}\left({\vec k}_1\cdot{\vec h}
(x^\prime_1,y_1) +{\vec k}_2\cdot{\vec h}(x^\prime_2,y_2)\right)}
={\rm e}^{-{\vec k}_1\cdot
{\vec k}_2G_0(x^\prime_1-x^\prime_2,y_1-y_2)}
:{\rm e}^{-{\rm i}({\vec k}_1
+{\vec k_2})\cdot{\vec h}(x^\prime_0,y_0)}\{1+\ldots\}:}
where $x^\prime_0=(x^\prime_1+x^\prime_2)/2$.
The MOPE is obtained by integrating over ${\vec k}_1$
and ${\vec k}_2$. More precisely, we define 
${\vec k}={\vec k}_1+{\vec k}_2$ and 
${\vec q}=({\vec k}_1-{\vec k}_2)/2$, so that 
${\vec k}_1\cdot{\vec k}_2=-q^2+{\cal O}({\vec k}^2)$.
This latter ${\cal O}({\vec k}^2)$ term can be set to zero 
if we are interested in the leading singularity, which is 
responsible for the divergence.
Integrating over ${\vec k}$ reconstructs a bi-local
$\delta$ operator and we get the following MOPE:
\eqn\MOPEtwo{\eqalign{
\delta^{(d-1)}&\left({\vec h}(x_1,y_1)-{\vec h}(x^\prime_1,y_1)\right)
\delta^{(d-1)}\left({\vec h}(x_2,y_2)-{\vec h}(x^\prime_2,y_2)\right)
\cr & = {1\over (4\pi)^{{d-1\over 2}}}
{1\over \left(-G_0(x_1-x_2,y_1-y_1)-G_0(x^\prime_1-x^\prime_2,y_1-y_2)
\right)^{{d-1\over 2}}} \cr &\quad \quad \quad \quad \quad \quad \quad 
\quad \quad \quad \quad \quad \times \delta^{(d-1)}
\left({\vec h}(x_0,y_0)-{\vec h}
(x^\prime_0,y_0)\right)+\ldots \ .\cr}}
We are interested in the pole in $\epsilon$ obtained when integrating
the coefficient of the $\delta$ term in the right hand side of 
\MOPEtwo\ over the relative coordinates $x_1-x_2$, 
$x^\prime_1-x^\prime_2$ and  $y_1-y_2$. Defining 
$Y=\vert y_1-y_2\vert+\vert x_1-x_2\vert^{1/2}+\vert x^\prime_1-x^\prime_2
\vert^{1/2}$, $u=\vert x_1-x_2\vert^{1/2}/\vert y_1-y_2\vert$
and $v=\vert x^\prime_1-x^\prime_2 \vert^{1/2}/\vert y_1-y_2\vert$,
and using again the explicit formula \propbis\ for $G_0$ and
$d=11-2\epsilon$, we obtain a pole in $\epsilon$ equal to:
\eqn\pole{32 {(2\sqrt{\pi})^5\over (4\pi)^5} \int_0^{\mu^{-1}}
dY {Y^4\over Y^{5-\epsilon}}\times \int_0^\infty du \int_0^\infty dv
{uv\over \left( f(u)+f(v)\right)^5}\ ,}
where $f(u)=u\, \exp (-1/4u^2) + (\sqrt{\pi}/2)\, {\rm erf}(1/2u)$.
The integral over $Y$ gives a pole $\mu^{-\epsilon}/\epsilon$.
The integral over $u$ and $v$ is convergent and will be denoted by:
\eqn\defI{I\equiv \int_0^\infty du \int_0^\infty dv
{uv\over \left( f(u)+f(v)\right)^5}={1\over 24}
\int_0^\infty da\, (F(a))^2 \ ,} 
where $F(a)\equiv a^2 \int_0^\infty du \, u\,{\rm e}^{-a\, f(u)}$.
The function $F(a)$ satisfies $F(a) \buildrel {a \to 0} \over 
\longrightarrow 1$ and $F(a) \buildrel {a \to \infty} \over \sim \exp
(- a \sqrt{\pi}/2)$.
The integral $I$ can be estimated numerically to $I=0.068373636(1)$.

Applying as before the MOPE of Eq.~\MOPEtwo\ to the {\it renormalized} 
theory, now expanded as in \Zpertexp\ to second order in $b^R$, 
the $N=2$ diagram gives two $\delta$ interactions with a factor 
$(b^R \mu^\epsilon)^2/8$, leading to a divergence equal to 
\eqn\diverg{2\times {(b^R\mu^\epsilon)^2 \over 8}{1\over \pi^{{5\over 2}}}
{\mu^{-\epsilon} I\over \epsilon} } 
with a factor of $2$ coming from the two ways of assembling the two
dipoles of the diagram into a one-loop molecule.
This divergence will be cancelled by the $\delta$ interaction counterterm 
in the renormalized Hamiltonian, which comes in the expansion with 
a factor $-(Z_b-1)b^R\mu^\epsilon/2$, provided
\eqn\zb{Z_b=1+ {b^R\over 2\pi^{{5\over 2}}} {I \over \epsilon}\ .}
Using \Zperpfinal\ and \zb , we relate the bare and renormalized
coupling constants as in \Rquant\ for $D=2$ and $d=11-2\epsilon$:
\eqn\bbr{b=\mu^\epsilon b^R \left(1+{I\over 2\pi^{{5\over 2}}}
{b^R\over \epsilon}\right)\left(1+{1\over 16\pi^2}{b^R\over \epsilon}
\right)^{7-\epsilon\over 2}
+{\cal O}\left((b^R)^3\right)\, }
leading, after differentiation with respect to $\mu$ at fixed $b$
to the one-loop Wilson function:
\eqn\difer{\beta(b^R)\equiv \mu\left.
{d\phantom{\mu}\over d\mu}\right\vert_0 b^R=-\epsilon b^R +
\left({I\over 2\pi^{{5\over 2}}}+{7-\epsilon\over 2}
{1\over 16\pi^2}\right)(b^R)^2+{\cal O}\left((b^R)^3\right)
.}
We thus obtain an infrared stable fixed point at
\eqn\fixedpoint{b^{R\star}={\epsilon\over 
{I\over 2\pi^{{5\over 2}}}+{7\over 2}{1\over 16\pi^2}}+{\cal O}(\epsilon^2)
\ .}
This fixes the value of the anomalous dimension $\delta$ through
\eqn\anomalous{\eqalign{\delta(b^R)\equiv
\mu\left.{d\phantom{\mu}\over d\mu}\right\vert_0 \log Z_\perp
=\beta(b^R)
{d\phantom{b^R}\over db^R} \log Z_\perp
&=\Big(-\epsilon b^R +{\cal O}\left((b^R)^2\right)\Big)
\left({1\over 16\pi^2}{1\over \epsilon}+{\cal O}\left(b^R\right)
\right)\cr &
=-{b^R\over 16 \pi^2} +{\cal O}\left((b^R)^2\right)\cr }
}
and 
\eqn\anom{\delta\equiv\delta(b^{R\star})=-{\epsilon\over
{8I\over \sqrt{\pi}}+{7\over 2}}+{\cal O}(\epsilon^2)\ .}
Numerical values for the exponents at $D=2$ and $d=3$ are obtained
by setting $\epsilon=4$ in the above formula, giving
\eqn\deltanum{\delta=-1.050
}
and thus the estimates
\eqn\numexp{\eqalign{z&=0.678
\cr
\nu&=0.517
\cr
\zeta&=0.762
\ .\cr}}
To understand the values we obtain for these exponents 
more clearly notice that the factor $7/2$ in
the denominator of \anom\ is actually the factor
$(d+3)/4=(7/2)-(\epsilon/2)$, appearing in the exponent 
of $Z_\perp$ in \Rquant\ , to first order in $\epsilon$. 
It is therefore legitimate, at first order, 
to replace this factor $7/2$ by the factor $3/2$ obtained 
by setting $d=3$ directly. This in practice amounts to making a 
partial two-loop correction.
This leads to a new estimate $\delta=-2.212$ and $\zeta=1.053$, well above
the original estimate \numexp , and actually even unphysical since larger 
than one. We see here that, due to the large value of $\epsilon=4$ at the 
physical dimension $d=3$, the first order estimates \numexp\ are not robust
with respect to second order corrections and cannot be reliable.

It is also interesting to develop alternative expressions for 
the roughness exponent $\zeta$, as was done for the isotropic
membrane Edwards model in \DW .
Indeed, the above estimate of $\zeta$ relies on the expression 
\expo\ expressing the deviation of $\zeta$ from its Gaussian
value $\zeta_0$ at $\epsilon=0$, in terms of the anomalous exponent
$\delta=\mu\left.{d\phantom{\mu}\over d\mu}\right\vert_0 \log Z_\perp$,
which we estimated to first order in $\epsilon$ in \anom .
Using the relation \Rquant\ between the bare and renormalized 
coupling constants, however, we can write, at the fixed point, the two
following equivalent definitions of $\delta$:
\eqn\deltafv{\eqalign{\delta &= \delta_F -{4 \over {\{ 4 + (D-1)(d+3)\}}}
\mu\left.{d\phantom{\mu}\over d\mu}\right\vert_0 \log {Z_b\over Z_\perp}\cr
&= \delta_{\rm var} -{4 \over (D-1)(d+3)}
\mu\left.{d\phantom{\mu}\over d\mu}\right\vert_0 \log Z_b 
\ ,\cr}}
leading directly to the two identities
\eqn\zetafv{\eqalign{\zeta &= \zeta_F +{D-1\over {\{ 4 + (D-1)(d+3)\}}}
\mu\left.{d\phantom{\mu}\over d\mu}\right\vert_0 \log {Z_b\over Z_\perp}\cr
&= \zeta_{\rm var} +{1 \over d+3}
\mu\left.{d\phantom{\mu}\over d\mu}\right\vert_0 \log Z_b 
\ .\cr}}
These relations express the deviation of $\zeta$ from its Flory
value and its Variational value respectively.
As we did for $\delta$ in \anomalous\ and \anom , we can obtain for 
$D=2$ and $d=11-2\epsilon$ the estimates to first order in $\epsilon$:
\eqn\foe{\eqalign{\mu\left.{d\phantom{\mu}\over d\mu}\right\vert_0 
\log {Z_b\over Z_\perp} &= {1-{8I\over \sqrt{\pi}} \over
{8I\over \sqrt{\pi}}+{7\over 2}}\epsilon +{\cal O}(\epsilon^2)
\cr
\mu\left.{d\phantom{\mu}\over d\mu}\right\vert_0 \log Z_b &=
=-{{8I\over \sqrt{\pi}} \over
{8I\over \sqrt{\pi}}+{7\over 2}}\epsilon +{\cal O}(\epsilon^2)
\ .\cr}}
One can easily check that 
the two equations in \zetafv\ give exactly the same estimate
as before for $\zeta$ at first order in $\epsilon$, provided that the 
quantities $\zeta_F$, $\zeta_{\rm var}$ and the different factors 
appearing in \zetafv , which involve $d$, are themselves expanded 
to first order in $\epsilon$.
 
On the other hand, one could also decide not to expand any of
these factors and impose $d=3$ directly. If one moreover restores
the factor $(d+3)/4$ instead of $7/2$, as discussed above, all the
various expressions reproduce the unphysical estimate 
$\zeta=1.053$. If only some of the terms are 
expanded in $\epsilon$, we get lower values of $\zeta$. 
We thus expect that the original estimate $\zeta=0.762$, obtained 
by expanding all terms at first order in $\epsilon$ is actually a 
lower bound on the exact value of $\zeta$.

\newsec{Conclusions}

In this paper we have studied, within the $\epsilon$ expansion,
the effects of self-avoidance in the tubule model introduced by RT,
going beyond their variational and Flory theory treatments of
self-avoidance.
We first show that the model is renormalizable 
and, furthermore, that the bending energy term is not renormalized.
We then derive very general scaling relations for the critical
exponents of the model at an infrared stable fixed point. 
These relations imply there is only one independent exponent.
For special choices of the renormalization factors we are
able to reproduce three different limits of the model viz.\
the trivial Gaussian model, the Flory approximation and the
Gaussian variational approximation \RT. This shows the power of 
this approach. We then treat the fluctuations of the model
to one-loop in the self-avoiding parameter in an $\epsilon$
expansion about the upper critical dimension. This yields
predictions for all the critical exponents to
first order in $\epsilon$.

One should notice that our results have been obtained for
an infinitely large membrane. For a finite membrane with
extension $L_y$ in the $y$-direction and $L_\perp$ in
the transverse direction, finite size scaling
laws can be derived in the above renormalization group framework \RT.
Due to the anisotropic nature of the tubular phase, however,
there are many different scaling regimes, depending
in particular on the relative scaling of $L_y$ and $L_\perp$.

Finally, let us stress that the above analysis of renormalizability
does not depend on the precise form of the Gaussian elastic term in 
the $y$-direction. One could imagine replacing the bending energy term
by a tension term $(\partial_y{\vec h})^2$, describing for instance
a tubule under longitudinal tension. The theory would then also be 
renormalizable in an $\epsilon=2D-1-(d-1)(2-D)/2$ expansion, with 
again no renormalization of this tension term and only one independent
exponent in the theory. 
In this case, however, the calculation cannot be performed
at $D=2$ directly where the upper critical dimension is infinite.
As for self-avoiding isotropic membranes, a complete study of the problem
for $D<2$ is required. 

After this paper was completed we were informed by RT that the  
Hamiltonian Eq.~\hamilt\ is not sufficient for a 
complete description of polymerized tubules in d=3.
RT argued that a more involved Hamiltonian, 
including the anharmonic elastic terms of RT Eq.~(5), in addition to
the self-avoiding interaction, is needed. 
Since the present paper is rigorous and self-consistent we feel that
it nevertheless makes a vital contribution to our present understanding
of tubules. The analysis of the fuller model suggested by the remarks above
presents a very definite challenge {---} to our knowledge
there does not exist in the literature any proper renormalization
group treatment of a theory with {\it both} nonlinear elasticity
and two-body self-avoidance. 
\bigskip
 
\noindent{\bf Acknowledgments}

We thank Leo Radzihovsky for very helpful discussions.
The research of M.B. was supported by the Department of Energy, USA,
under contract N$^{\rm o}$ DE-FG02-85ER40237. We thank Fran\c{c}ois David 
and Marco Falcioni for a critical reading of the manuscript. 
 
\listrefs
 
\end